\documentclass[]{aa}
\usepackage{natbib}
\usepackage[dvips]{graphicx}
\usepackage{times}
\def\ebv{\mbox{$E_{B-V}$}}

\def\lya{\mbox{Ly$\alpha$}}
\def\deg{\hbox{$^\circ$}}
\def\ecs{\mbox{~erg~cm$^{-2}$~s$^{-1}$}}

\def\ecsa{\mbox{~erg~cm$^{-2}$~s$^{-1}$~{\AA}$^{-1}$}}
\def\lesssim{\mathrel{\hbox{\rlap{\hbox{\lower4pt\hbox{$\sim$}}}\hbox{$<$}}}}
\def\gtrsim{\mathrel{\hbox{\rlap{\hbox{\lower4pt\hbox{$\sim$}}}\hbox{$>$}}}}

\begin{document}

\title{Integral Field Spectroscopy of Extended \bf{\lya} Emission from the DLA
  Galaxy in Q2233+131}
\author{L.~Christensen\inst{1}
  \and S.~F.~S\'anchez\inst{1}
  \and K.~Jahnke\inst{1}
  \and T.~Becker\inst{1} 
  \and L.~Wisotzki\inst{1,2}
  \and A.~Kelz\inst{1} 
  \and L.~\v C.~Popovi\'c\inst{1,3}
  \and M.~M. Roth\inst{1}
}
\institute{Astrophysikalisches Institut Potsdam, An der Sternwarte 16, 
     14482 Potsdam, Germany
\and  Potsdam University, Am Neuen Palais 10, 14469 Potsdam, Germany
\and  Astronomical Observatory, Volgina 7, 11160 Belgrade 74, Serbia
}

\mail{lchristensen@aip.de}
\date{Received 22 September 2003 / Accepted 15 December 2003}

\titlerunning{Extended \lya\ emission from the DLA galaxy in Q2233+131}


\abstract{This paper presents observations of an extended Lyman-$\alpha$
  emission nebula surrounding the galaxy responsible for the Damped
  Lyman-$\alpha$ Absorption (DLA) line in the spectrum of the quasar
  \object{Q2233+131}. With the Potsdam Multi Aperture Spectrophotometer (PMAS)
  we measure the properties of the extended \lya\ emission in an area of
  3\arcsec$\times5$\arcsec\ having a total line flux of
  $(2.8\pm0.3)\times10^{-16}$\ecs, which at redshift $z=3.15$ corresponds to a
  luminosity of ($2.4_{-0.2}^{+0.3})\times10^{43}$~erg~s$^{-1}$ and a size of
  23$\times$38 kpc. The location of the emission is spatially coincident with
  the previously detected DLA galaxy, but extends significantly beyond its
  limb.  We argue that the \lya\ emission is likely to be caused by an outflow
  from the DLA galaxy, presumably powered by star formation.  In the case of
  negligible dust extinction, the \lya\ luminosity indicates a star-formation
  rate of $19\pm10$~M$_{\odot}$~yr$^{-1}$ consistent with that derived from
  the UV continuum flux from the parent galaxy. The wind velocity indicated by
  the integral field spectra is of the order of several hundred km s$^{-1}$.
  We find no indication of emission originating in a rotating disk.
   } 

  \maketitle

\section{Introduction}
\label{intro}
High redshift quasars (QSOs) show multiple absorption lines bluewards of the
redshifted 1216 {\AA} Lyman$\alpha$ (\lya) wavelength.  Known as the
Lyman$\alpha$ forest, this is caused by absorption in neutral hydrogen clouds
along the line of sight towards the QSO. Clouds having column densities larger
than 2$\times$10$^{20}$cm$^{-2}$ give rise to line profiles with broad wings
characteristic of damped Ly$\alpha$ lines. To date approximately 150 Damped
Ly$\alpha$ Absorbers (DLAs) with redshifts of $0.1<z<4.6$ have been confirmed
\citep{curran02}.  It has been found that DLAs contain a significant fraction
of total gas mass compared to the mass of the stars in present day galaxies
\citep{wolfe95,stolom96,stolom00}, and it is questioned whether there is a
significant evolution with redshift \citep{rao00}.  Spectroscopic observations
have shown that the DLAs have metallicities of 0.01--1 times solar with a mild
increase with decreasing redshift \citep{prochaska03}, suggesting that DLAs
are star-forming objects. Nevertheless, the relation between DLAs and galaxies
is not well understood.  At higher redshifts, the DLA galaxies have been
suggested to be thick disks or the progenitors of present day spirals galaxies
\citep{wolfe86}, while others suggest that the counterparts could be dwarfs
\citep{hunstead90}, or galaxy building blocks in a hierarchical merging
scenario \citep{haehnelt98}.

Many investigations have been performed in order to establish what the galaxy
counterparts to the DLAs resemble most
\citep[e.g.][]{lebrun97,warren01,colbert02}.  Typically, deep broad-band or
narrow-band imaging of the fields containing the DLAs has been carried out,
and objects near the line of sight of the QSOs are detected after subtraction
of the QSO point spread function. These candidate DLA galaxies are typically
faint. Successive follow-up spectroscopy of the candidates is required to
reveal whether or not they have the same redshift as the DLA line.  Only in 4
cases DLA galaxies have been confirmed this way for the high redshift
($z\gtrsim1.9$) DLA galaxies \citep{moller93,djorgovski96,fynbo99,moller02}.
In two additional cases \lya\ emission lines have been detected in the troughs
of the DLA lines in the QSO spectra \citep{leibundgut99,ellison02}.

We here present a study of the Q2233+131 at $z=3.295$, which has a DLA line at
$z=3.153$ \citep{sargent89}. The metallicity of the DLA is [Fe/H]~=~--1.4
\citep{lu97}, and the column density of H\,I is below the classical limit of a
DLA line, having $N_{\ion{H}{I}}=1\times10^{20}$~cm$^{-2}$, thus formally
  characterizing this as a Lyman-Limit system. In accordance with previous
  papers on this object we will continue to denote it a DLA absorber.  A
  candidate galaxy responsible for the absorption was found at an impact
  parameter of 2\farcs3 using the Lyman break technique, suggesting a redshift
  larger than 3 \citep{steidel95}.  This object was confirmed as the absorbing
  galaxy having the same redshift as the DLA line \citep[][hereafter
  D96]{djorgovski96}. These authors measured a \lya\ line flux of
  ($6.4\pm1.2)\times10^{-17}$\ecs\ in a Keck long-slit spectrum and found the
  magnitudes of the Lyman break galaxy of $R=24.8\pm0.1$ and $V=25.1\pm0.2$.
  \citet{warren01} found $H=25.34\pm0.17$ for the galaxy using near-IR
  photometry with the HST/NICMOS, and \citet{moller02} found
  $V_{50}=25.75\pm0.12$ with STIS images.
  
  With integral field spectroscopy the conventional two step approach for
  confirming a DLA galaxy can be avoided. A previous attempt to use this
  technique only yielded an upper limit for the line emission from the DLA
  galaxy in \object{BR 1202--0725} \citep{petitjean96}, while the spectral
  range of the observations of the DLA system in \object{APM~08279+5255} did
  not cover the appropriate wavelengths for the redshifted \lya\ emission
  \citep{ledoux98}.  We show here that with the Potsdam Multi Aperture
  Spectrophotometer (PMAS) instrument we not only detect \lya\ emission from
  the DLA galaxy at $z=3.15$, but we find that the object causing the \lya\ 
  emission is extended, and the line flux is larger than reported previously
  in the literature.

In Sect. \ref{data} of this paper we will describe the spectroscopic
observations, and the procedures for reducing the data. We analyse the spectra
of the QSO and the extended emission from the DLA in Sects.~\ref{qsospec} and
\ref{dlaspec}, respectively, addressing the nature of the extended emission.
The relation between the \lya\ emission and the location of the DLA galaxy
seen in high spatial resolution Hubble Space Telescope (HST) images is
described in Sect.~\ref{wfpc}. Similarities with other \lya\ emitting objects
are described in Sect.~\ref{hi_z}, which leads to possible interpretations for
the origin of the extended \lya\ nebula in Sect.~\ref{origin}.  The ionised
gas mass in the nebula is estimated in Sect.~\ref{mass}. In Sect.~\ref{conc}
we present our conclusions.

Throughout the paper we assume a flat Universe with $H_0=70$ km s$^{-1}$
Mpc$^{-1}$, $\Omega_m=0.3$, and $\Omega_{\Lambda}=0.7$. The redshift $z=3.15$
then corresponds to a luminosity distance of $8.3\times10^{28}$~cm (27.0~Gpc),
1\arcsec\ corresponds to a linear size of 7.6 kpc, and the look-back time is
11.5 Gyr.


\section{Observations and data reduction}
\label{data}
The PMAS integral field instrument uses two cameras: A cryogenic acquisition
and guiding camera (A\&G camera) that can be used for imaging in addition to
the integral field spectrograph (IFS) \citep{pmas00}. The A\&G camera has a
SITe TK1024 chip with 1k$\times$1k pixels with a scale of 0\farcs2 per pixel
giving a field of view of 3\farcm4$\times$3\farcm4. The PMAS spectrograph is
equipped with 256 fibers coupled to a 16$\times$16 lens array, that we used
with a spatial sampling of 0\farcs5$\times$0\farcs5 per fiber on the sky,
resulting in a field of view of 8\arcsec$\times$8\arcsec. The spectrograph
camera has a SITe ST002A 2k$\times$4k CCD and its 256 spectra have a FWHM of
$\sim$2 pixels when using a 2$\times$2 binned read-out mode. The spectra are
aligned on the CCD with 7 pixels between adjacent spectra making
cross-contamination negligible.

PMAS is mounted on the 3.5m telescope at Calar Alto. We observed Q2233+131 on
Sep. 2 2002 for a total of 7200~s (4$\times$1800~s) at an airmass between 1.094
and 1.15 and a seeing between 1\farcs0 and 1\farcs3 measured by the A\&G
camera. Using a 300 gr/mm grating yielded a spectral resolution of 6~{\AA}
while the grating was set to cover the wavelength range of 3930--7250 {\AA}.
Calibration images were obtained following the science exposures and consisted
of spectra of emission line lamps (HgNe), and spectra of a continuum lamp
needed to locate the 256 individual spectra on the CCD.  Observations of the
spectrophotometric standard stars \object{BD~+28\deg 4211} and Hz4 were
obtained during the night for flux calibration.

Reduction of the data was done in IDL with P3D\_online, a software package
written specifically for reducing PMAS data \citep{becker01}. After bias
subtraction the 256 spectra were extracted from the two-dimensional frames.
Wavelength calibration was performed using the emission line lamp spectra.
For the flat fielding the average transmission of each fiber was determined
using exposures of the sky at twilight. Cosmic ray hits were removed from each
of the 4 files using the L.A.Cosmic routine within IRAF \citep{vandok01}, and
the rejected pixels were inspected by eye checking that no pixels close to the
\lya\ emission line were affected. The data cubes were corrected for the
effect of differential atmospheric refraction using the formula of
\citet{filippenko02}. Given the small airmass of Q2233+131 during the
observation the effect is small, but not negligible at long wavelength
intervals. At the wavelength of the DLA line the effect can be ignored, since
this was the reference wavelength for estimating the relative offsets before
combining the 4 frames.  This method ensures that the position of the DLA
galaxy relative to the centroid of the QSO at other wavelengths is not
shifted.  In the end the four files were co-added, resulting in a data cube of
dimensions 16$\times$16$\times$1024 pixels.

For subtraction of the sky background, an average background spectrum was
created by co-adding several spectra at the edge of the field of view,
uncontaminated by the QSO flux, and subtracted from all 256 individual
spectra. All further data calibration was done using IRAF.  For the standard
star observations we coadded all spectra within a radial aperture of 3\arcsec,
and compared the one-dimensional standard star spectrum with table values to
create a sensitivity function taking into account the atmospheric extinction
typical for Calar Alto \citep{hopp02}. Finally, the spectra were flux
calibrated using this sensitivity function. 

Further analysis and inspection of the data was performed with the ``Euro3D
Visualization Tool'', which is a very efficient tool made for visualizing
integral field data \citep{sanchez03}.

\section{The QSO spectrum}
\label{qsospec}

\begin{figure*}
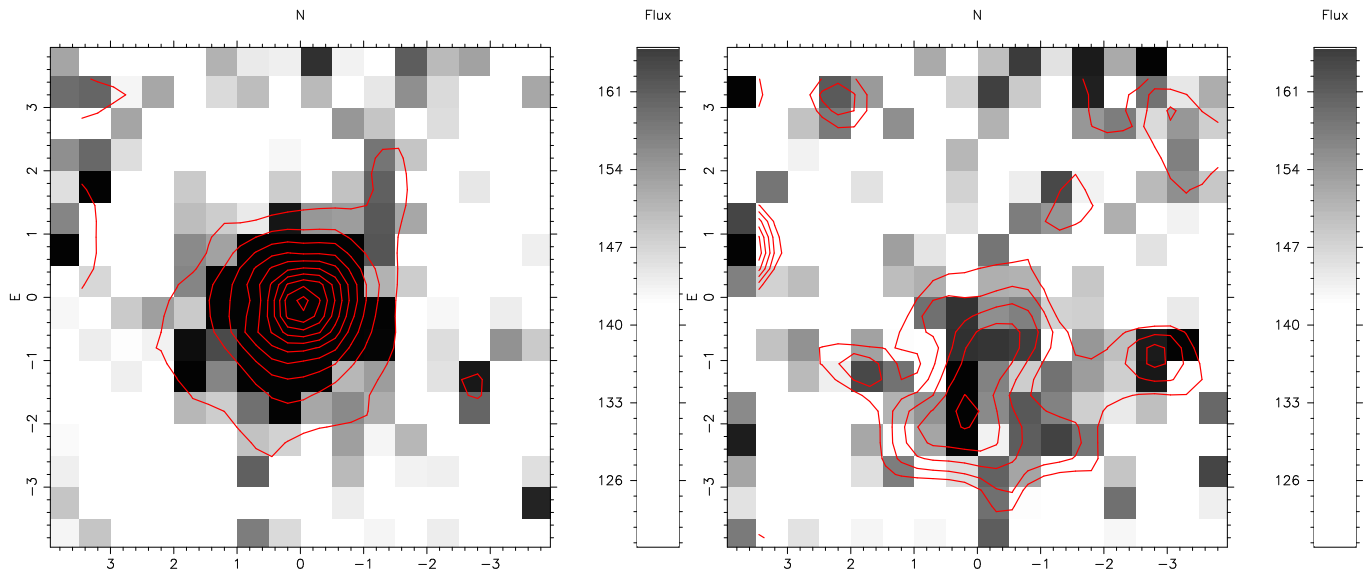

\begin{minipage}[c]{0.5\textwidth}
\resizebox{\hsize}{!}{\includegraphics[angle=-90]{qso_map.ps}}
\end{minipage}%
\begin{minipage}[c]{0.5\textwidth}
\resizebox{\hsize}{!}{\includegraphics[angle=-90]{source_map.ps}}
\end{minipage}
\caption{Left hand panel: Image of Q2233+131 in the wavelength range 
  5200--5300~{\AA}. This image shows the 16$\times$16 spatial pixel
  (``spaxel'') field of view of PMAS corresponding to
  8\arcsec$\times8\arcsec$. The QSO is centered in the field at (0,0).  Right
  hand panel: Co added narrow band image of the Q2233+131, with a field of
  view of 8\arcsec$\times8\arcsec$, north is up and east is left.  Selecting
  the wavelength range 5040--5055~{\AA} an extended object of roughly
  5\arcsec$\times3\arcsec$ appears, which is the extended \lya\ emission from
  the DLA galaxy. The selected wavelengths correspond to the DLA absorption
  trough in the QSO spectrum.  The centroid of the QSO is at (0,0) while the
  position of the DLA galaxy found from the broad band observations in D96 and
  in the space-based images is at (--0\farcs9,--2\farcs2).  Apparently, the
  extended source overlaps with the position of the QSO, but this is caused by
  the fact that some QSO emission is present at the longest selected
  wavelengths ($\lambda > 5050${\AA}), as emission from the red wing of the
  DLA line has been included.  Contours of 5--8$\sigma$ levels above the
  background have been overplotted for guiding the eye.  The contours
  correspond to the combined emission from the \lya\ nebula and the QSO. 
    The high signal at the left edge of the field at (3.5,1) is caused by a
    bad fiber.}
\label{fig:map}
\end{figure*}

\begin{figure*}
  \hspace{-1cm}\resizebox{19cm}{!}{\includegraphics[angle=90]{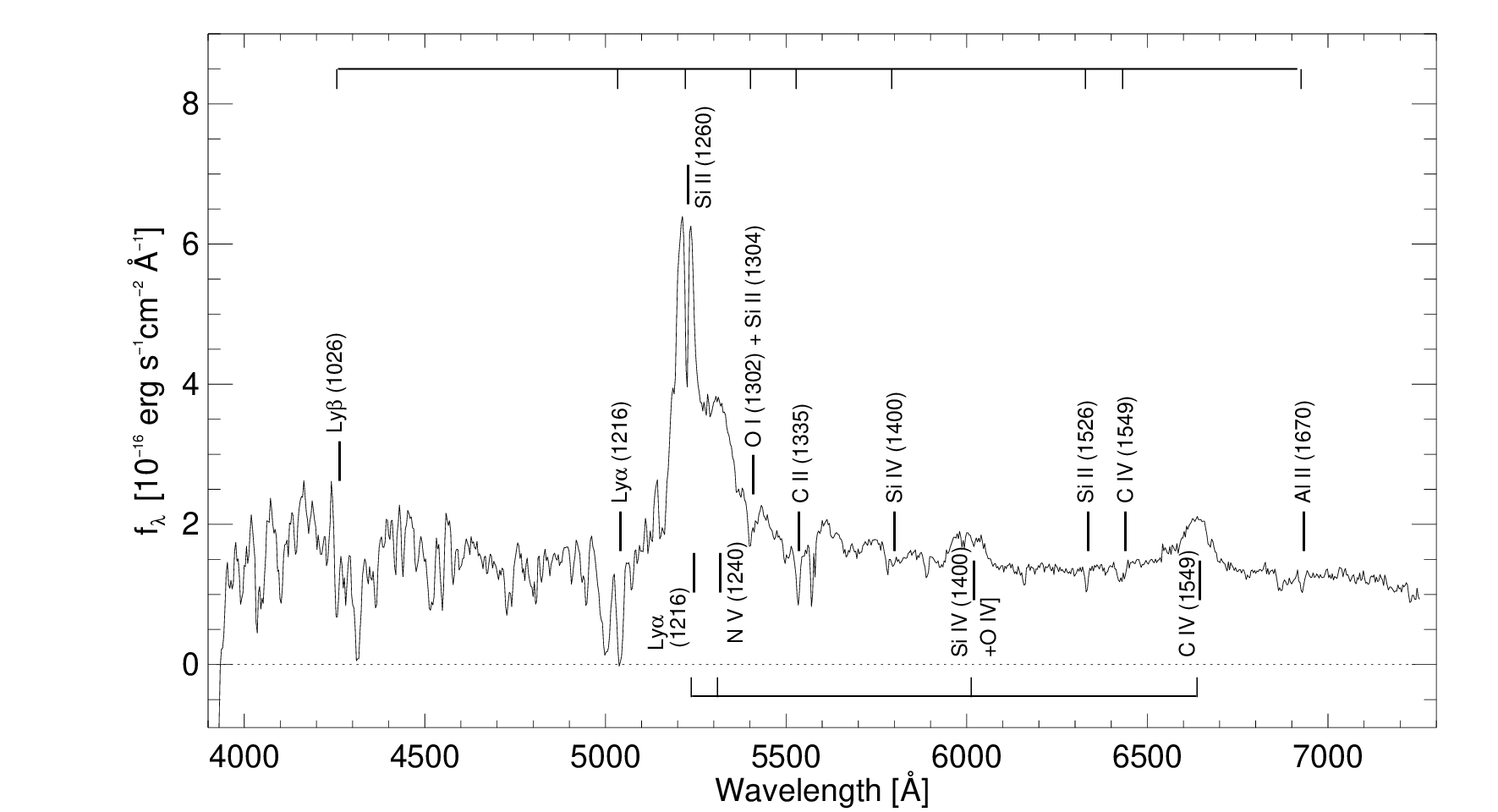}}
\caption{Spectrum of  Q2233+131. All spaxels within a radial
  aperture of 2\arcsec\ have been coadded and then scaled to the total flux of
  the QSO, measured within a 3\arcsec\ radial aperture. The bright line at
  5200~{\AA} is \lya\ emission from the QSO. This and other emission lines
  from the QSO are labeled under the spectrum. The DLA line at 5050 {\AA} and
  the corresponding metal absorption lines belonging to this system are
  labeled above the spectrum. Redshifts of all lines are given in
  Table~\ref{tab:lines}.}
\label{fig:spec}
\end{figure*}

We constructed narrow band images by selecting appropriate wavelength
intervals in the combined 3D data cube, resulting in images with 16$\times$16
spatial pixels (``spaxels''). In Fig.~\ref{fig:map} such an image of Q2233+131
is shown in the wavelength range 5200--5300~{\AA}. For creating a
one-dimensional object spectrum one can select spaxels in this image, each of
which represents a single spectrum, and co-add the selected spectra.

For comparison with later spectra we created a  combined spectrum of Q2233+131
by co-adding all spectra within a radial aperture of 2\arcsec\ ($\sim$45
spaxels) in Fig.~\ref{fig:map}. An overall aperture correction of 2\% was
applied, and the resulting spectrum is shown in Fig.~\ref{fig:spec}.

We detect many absorption features from metal lines in the spectrum of the QSO
associated with the $z=3.15$ DLA system, some of which were already recognized
in D96. The redshifts of these lines, listed in Table~\ref{tab:lines}, matches
the redshift of the DLA line. We derive a mean systemic redshift
$z=3.1475\pm0.0005$ from the all metal lines in the DLA apart from the
\ion{Si}{IV} and \ion{O}{IV} 1400 {\AA} blend which has a larger offset than
the low ionization species.  A detailed analysis of the metallicity and column
densities is outside the scope of this paper, as higher spectral resolution
would be necessary for this purpose. For the same reason the Ly$\beta$
absorption line is blended with the Lyman$\alpha$ forest. The dip seen in the
tip on the \lya\ emission from the QSO is probably caused by \ion{Si}{II}
1260~{\AA} at the redshift of the DLA system.

In addition to the absorption lines we find broad emission lines from \lya,
\ion{N}{V} 1240, \ion{Si}{IV} + \ion{O}{IV} 1400 {\AA} and \ion{C}{IV} 1549
{\AA} from the QSO.  Including these four lines we find the redshift
$z=3.2877\pm0.0052$ for the QSO.

\begin{table}
\centering
\begin{tabular}{llll}
\hline \hline  
\noalign{\smallskip}
Ion            & $\lambda_{\mathrm{lab}}$  [{\AA}] & $z_{\mathrm{em}}$   & $z_{\mathrm{abs}}$ \\
\noalign{\smallskip}
\hline
\noalign{\smallskip}
\lya                                 & 1215.67  & 3.2949 \\
\ion{N}{V}                           & 1238.82, 1242.80 & 3.2840 \\
\ion{Si}{IV} + \ion{O}{IV]} (blends) & 1400     & 3.2881 \\
\ion{C}{IV}                          & 1548.20, 1550.78 & 3.2838 \\
\lya                                 & 1215.67  & &3.1476 \\
\ion{Si}{II}                         & 1260.42  & &3.1468 \\
\ion{O}{I}                           & 1302.16  & &3.1479 \\
\ion{C}{II}                          & 1334.53  & &3.1476 \\
\ion{Si}{IV}                         & 1400     & &3.1490\\
\ion{Si}{II}                         & 1526.71  & &3.1485 \\
\ion{C}{IV}                          & 1548.20, 1550.78 & &3.1477\\
\ion{Al}{II}                         & 1670.789 & &3.1478 \\
\noalign{\smallskip}
\hline
\noalign{\smallskip}
\lya (DLA galaxy) &                    1215.67 & 3.1538 \\
\hline
\noalign{\smallskip}
\end{tabular}
\caption[]{List of emission lines from the Q2233+131 itself and the 
 absorption lines, related to the $z~=~3.15$ DLA system detected in the 
 QSO spectrum. A standard air-to-vacuum correction has been applied to
 the observed lines before deriving the redshifts. The uncertainties of 
 the measured redshifts are typically  $\pm$0.0005.}
\label{tab:lines}
\end{table}

\section{The spectrum of the DLA galaxy}
\label{dlaspec}

\begin{figure}
\hspace{-0.8cm}\resizebox{9.8cm}{!}{\includegraphics[angle=90]{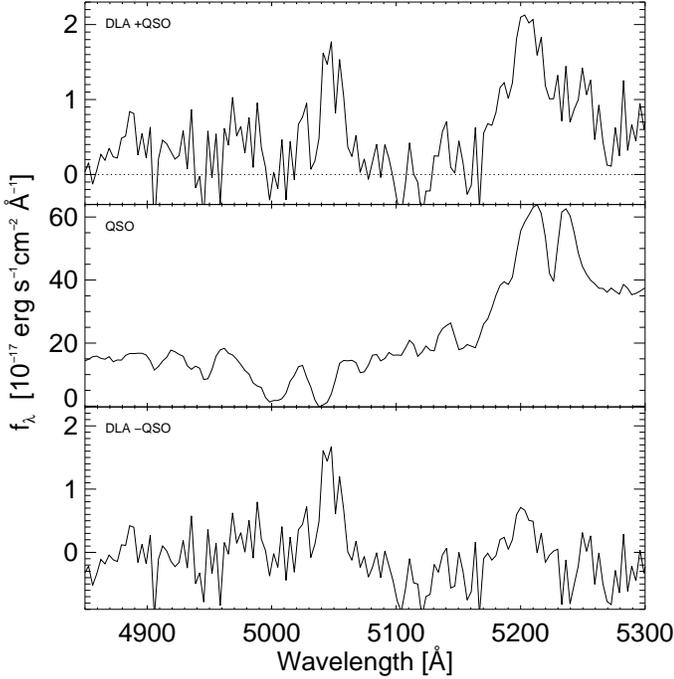}}
\caption{
  Spectrum of the extended emission associated with the DLA galaxy at $z=3.15$
  in Q2233+131. The spectrum has been obtained by co-adding $\sim$35 spectra
  which appear to be associated with the DLA galaxy in the image in
  Fig.~\ref{fig:map}.  The few spaxels which are overlapping the position of
  the QSO have not been coadded, since these are contaminated by flux from the
  wing of the QSO and not associated with the \lya\ emission object. An
  emission line at 5050~{\AA} corresponds to \lya\ at the redshift of the DLA
  line in the QSO. The broad emission feature at $\sim$5200~{\AA} is due to
  residual \lya\ emission from the QSO at a radial distance of 2\arcsec\ from
  the centre of the QSO.  For comparison we have plotted the total QSO
  spectrum in the middle panel with the same wavelength range.  Note the scale
  on the y-axis differs by a factor of $\sim$30 between the two plots. In the
  lower panel we have plotted the DLA spectrum where a scaled QSO spectrum has
  been subtracted.  }
\label{fig:spec_dla}
\end{figure}

In Fig.~\ref{fig:map} we show another narrow-band image, this time selected
from the wavelength interval 5040--5055 {\AA}, corresponding to the absorption
trough in the QSO spectrum. An extended object is visible to the south and
south-west of the location of the QSO. The size of this object is roughly
5\arcsec$\times3\arcsec$, however, some of the extended object emission is due
to the presence of emission in the red wing of the DLA line from the QSO.  We
note that the very high signal which appears  at the left edge of the
  field with coordinates (3.5,1) is due to a bad flat-field effect of one
single fiber, which has a lower overall transmission.

We have added 35 spaxels which are apparently associated with the emission
line object. A part of the spectrum around the emission line is shown in the
upper panel in Fig.~\ref{fig:spec_dla} where one clearly sees an emission line
at $\sim$5050~{\AA}.  This wavelength corresponds to \lya\ at the redshift of
the DLA line at $z=3.1538\pm 0.0005$, in agreement within 1$\sigma$ with the
redshift published in D96. A second line is present at $\sim$5200~{\AA} which
is caused by the \lya\ emission from the QSO at a distance of 2\arcsec.  For
comparison the same section of the QSO spectrum is shown in the middle panel
in Fig.~\ref{fig:spec_dla}. There appears to be another broad absorption line
at 5000~{\AA}, but a high resolution spectrum of the QSO has shown that this
feature is caused by a blend of 5--6 individual absorption lines
\citep{bechtold94}. The lower panel shows the spectrum of the DLA where a
scaled spectrum of the QSO has been subtracted. In fact, the \lya\ emission
line at 5050 {\AA} is effected very little by this subtraction.

The spectra were analysed with the ONEDSPEC package in IRAF.  We measured a
full width half maximum (FWHM) of the \lya\ emission line of $20\pm2${\AA}.
The resolution in the combined spectra is 7.9~{\AA}, measured from the FWHM of
the 5577~{\AA} night sky line in the combined spectra. This yields an internal
FWHM of $18.4\pm3.3$ {\AA} of the \lya\ line, corresponding to a restframe
velocity of $1090\pm190$~km~s$^{-1}$.  We see no emission line from
\ion{N}{V} $\lambda$1240 or \ion{C}{IV} $\lambda$1549 from the DLA galaxy
down to a 3$\sigma$ detection limit of 1$\times10^{-17}$\ecs\ in agreement
with D96, albeit their detection limit was 10 times fainter.

The redshift difference between the DLA galaxy's \lya\ emission component and
the DLA absorption components in the QSO spectrum is $\Delta
z=0.0063\pm0.0007$ corresponding to a velocity of 450$\pm$50~km~s$^{-1}$.
This is larger than the 209~km~s$^{-1}$ reported in D96. Here we have used
information from all the detected absorption lines in calculating the systemic
redshift of the DLA.  If we instead used the redshift reported in D96 for the
absorption minimum of the DLA line itself, we would find a velocity difference
of 270$\pm$40~km~s$^{-1}$, which is within 2$\sigma$ of their value. Since the
interstellar absorption lines have been shown to be blue-shifted by
$>-100$~km~s$^{-1}$ with respect to the stars in Lyman Break Galaxies
\citep{shapley03}, the latter value ($\sim$300~km~s$^{-1}$) is probably a
better estimate.

The line flux measured by fitting the observed line with a Gaussian profile is
$(2.4\pm0.3)\times10^{-16}$~\ecs; a simple summation of the flux values for
each pixel from the emission line gives the same result within the errors.
From the dust maps of \citet{schlegel98} a Galactic reddening of $\ebv=0.068$
in the direction towards Q2233+131 is found. Correcting for this effect
increases the line flux to ($2.8\pm0.3)\times10^{-16}$ \ecs\ and the total
luminosity of the source is $2.4_{-0.2}^{+0.3}\times10^{43}$ erg~s$^{-1}$ in
the adopted cosmology.  With the noise in each spectral element around
5000~{\AA} of 3.2$\times10^{-18}$~\ecsa\ and the line flux being measured over
20 pixels, the significance of the detection of the \lya\ line is 15$\sigma$.
The size of the object is 23$\times$38~kpc above a 3$\sigma$ detection
threshold of $1\times10^{-17}$ \ecsa.

A measurement of the equivalent width (EW) of the emission line is hampered by
the fact that the continuum emission is extremely faint. From the observed
ground-based $V$ and $R$ band magnitudes in D96 together with magnitudes from
HST images (derived in Section~\ref{wfpc}), we estimate the flux from the
underlying continuum at $\sim$5050~{\AA} to be
$(1.1\pm0.4)\times10^{-19}$~\ecsa\ assuming a power-law spectral energy
distribution, \(f_{\nu}\propto\nu^{\beta}\) in the continuum.  With this line
and continuum flux we derive the rest frame EW=$190_{-70}^{+150}$~{\AA}. The
large error is mainly caused by the uncertainty for the continuum flux.
Furthermore, the measured EW must be taken as a lower limit because the effect
of dust extinction affects the \lya\ flux more than the UV continuum emission.

\subsection{Velocity structure}
\begin{figure*}
\begin{minipage}[c]{.5\textwidth}
\resizebox{\hsize}{!}{\includegraphics[bb=75 230 752 800,clip,angle=-90]{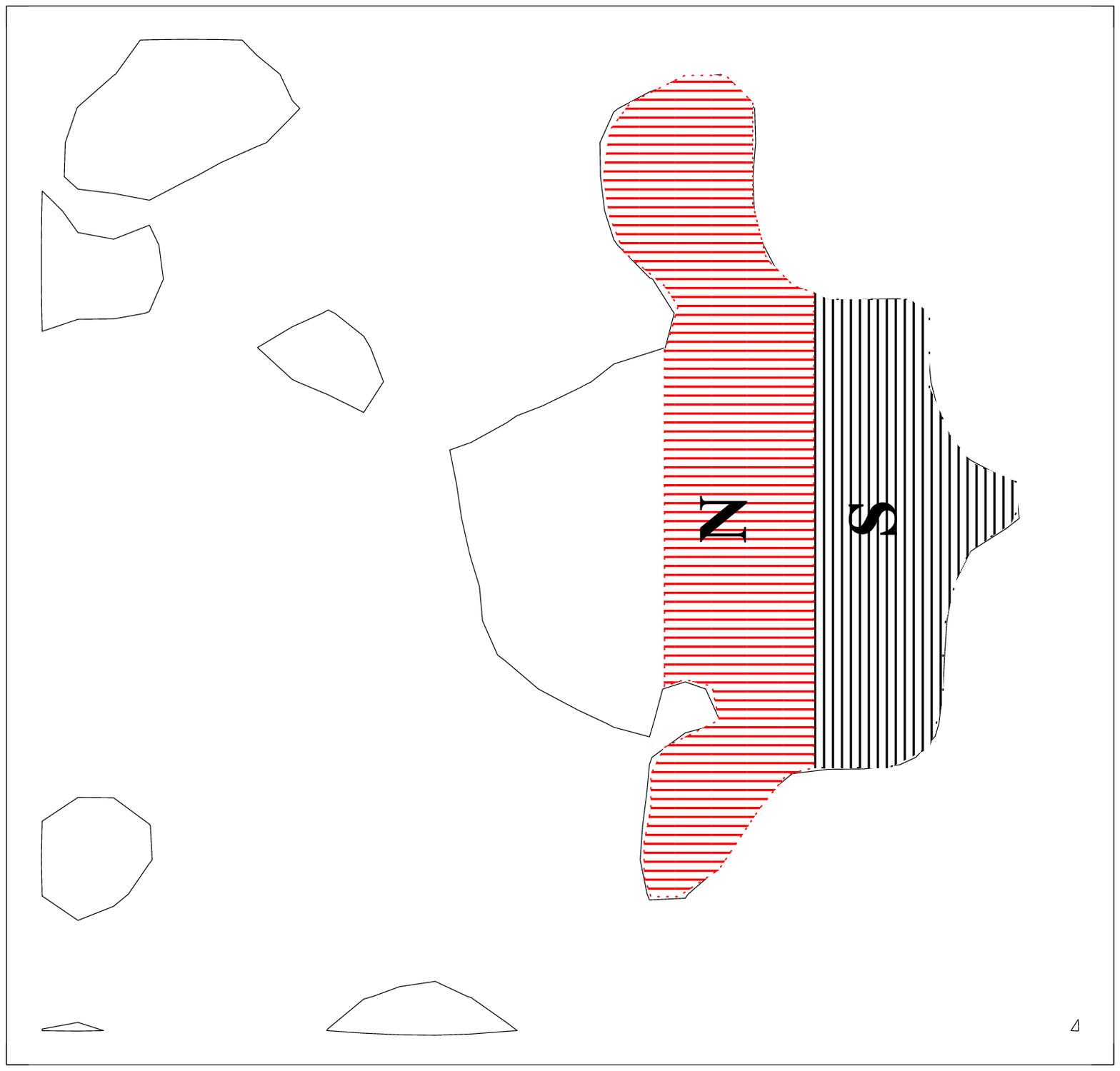}}
\end{minipage}%
\hspace*{-1cm}
\begin{minipage}[c]{.5\textwidth}
\resizebox{\hsize}{!}{\includegraphics[bb=75 230 685 715,clip,angle=-90]{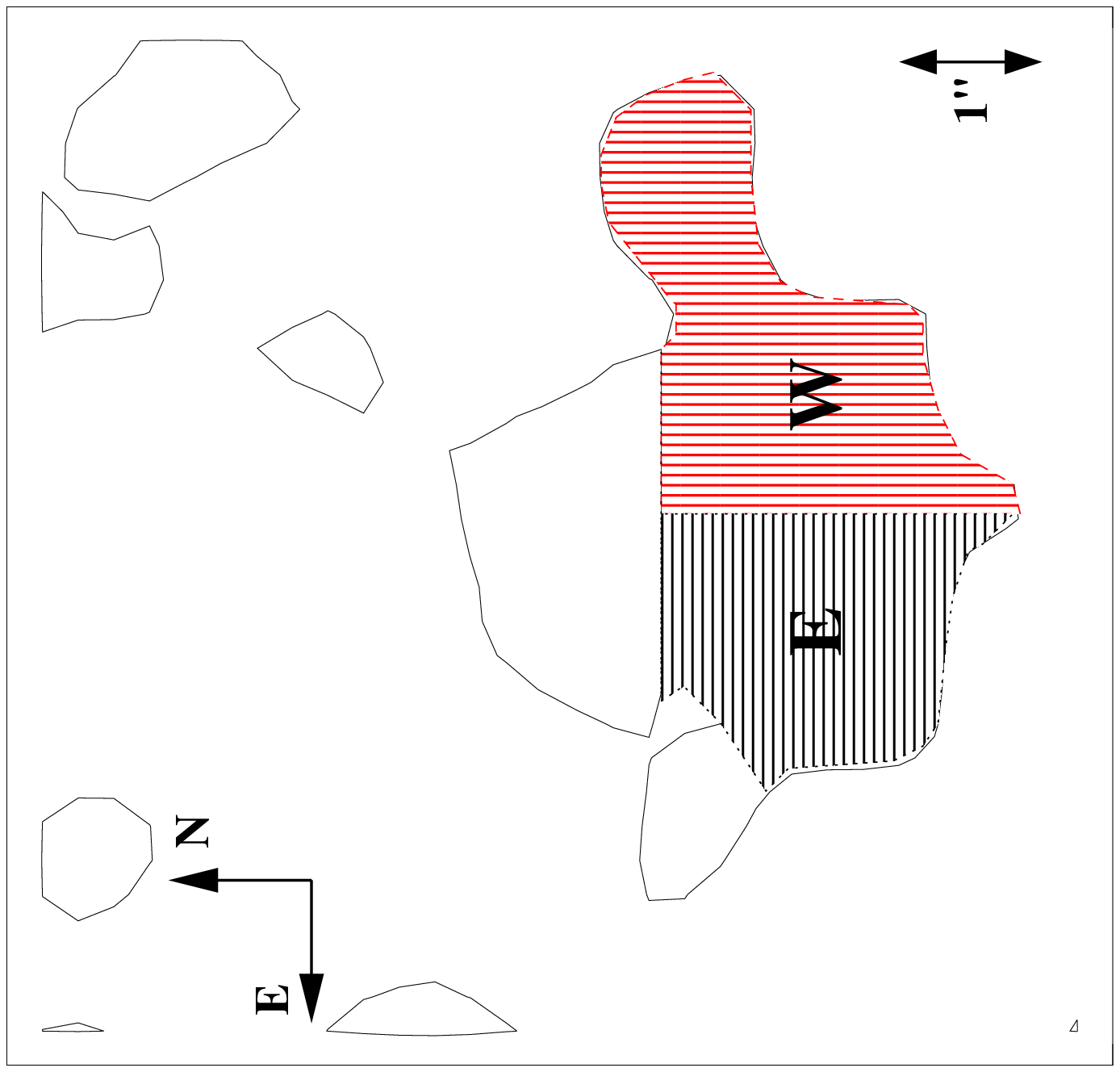}}
\end{minipage}
\vspace*{-3cm}
\caption{The images show the lowest (5$\sigma$) contour level taken from
  Fig.~\ref{fig:map}.  The chosen division between a north and a south region
  is indicated in the left panel, and the division between an east and a west
  region to the right. In both cases the emission spatially coincident with
  the QSO has not been included.}
\label{fig:nsew}
\end{figure*}

With integral field spectra one can in principle determine the velocity
structure of the \lya\ nebula. Specifically, it would be interesting to search
for signs of rotation. A detailed analysis of the velocity structure is not
possible with the present dataset, since in several of the individual spectra
the emission line is only detected on a $\sim$2$\sigma$ level, which makes any
analysis of any velocity structure unreliable.  Instead we summed up the
$\sim$15 spectra in the left-most (eastern) part and the $\sim$15 in the
right-most (western) part, respectively. A sketch of the division is shown in
Fig.~\ref{fig:nsew}. The \lya\ emission for each region is still detectable,
and the line flux of each is $\sim$1$\times10^{-16}$~\ecsa.  We fitted
Gaussians to the emission line in these two regions, and find that the peak of
the emission is shifted by +2.5 {\AA} from the east to the west part, which
corresponds to a rest-frame velocity difference of $\sim$150~km~s$^{-1}$.
This is small compared to the rotational velocity of present day large spiral
galaxies.  Repeating this exercise by splitting the extended object into a
southern and a northern part as shown in Fig.~\ref{fig:nsew}, the emission
line is shifted by 5 {\AA} corresponding to a difference of
$\sim$300~km~s$^{-1}$, where the northern part has the largest redshift.  This
velocity is comparable to that expected for spiral galaxies, but the geometry
is unusual for a disk, and as shown in Sect.~\ref{double} the apparent shift
of the emission line can be caused by a combination of more than one emission
region. In Table~\ref{tab:lya} we summarize the properties for the different
parts of the extended \lya\ nebula, and in Fig.~\ref{fig:sep} we show the
\lya\ line for the different regions. The  1$\sigma$ errors for the
centroids of the Gaussian profiles are 0.5 {\AA} estimated from simulations of
artificial spectra having faint emission lines at known wavelengths.  From
  these simulated spectra we also estimate that the 1$\sigma$ error for the
  EWs is $\sim$2 {\AA}.  All regions are seen to have a large FWHM indicating
velocities of 700--1100~km~s$^{-1}$, i.e. they are well resolved by our
spectra.

\begin{table}
\centering
\begin{tabular}{lllll}
\\  \hline \hline  
\noalign{\smallskip}
region & centroid & $f_{\mathrm{line}} $ & FWHM & EW \\
     &  ({\AA})     &  (\ecs)    &  (km s$^{-1}$) & ({\AA})\\
\noalign{\smallskip}
\hline
\noalign{\smallskip}
total   & 5048.3 & 2.4$\times10^{-16}$ & 1090$\pm$190 & $190_{-70}^{+150}$\\
east    & 5049.6 & 1.0$\times10^{-16}$ & 770$\pm$150    \\ 
west    & 5052.1 & 1.3$\times10^{-16}$ & 1120$\pm$200   \\
south   & 5048.8 & 0.7$\times10^{-16}$ & 650$\pm$150    \\
north   & 5053.4 & 1.3$\times10^{-16}$ & 890$\pm$150    \\
\noalign{\smallskip}
\hline
\noalign{\smallskip}
\end{tabular}
\caption[]{
Properties of the \lya\ emission from different parts of the extended
nebula derived from fitting a single Gaussian profile to the observed 
spectrum.  The  1$\sigma$ errors of the estimations of the Gaussian centroids
are 0.5~{\AA}. Note that there is an overlap of the selected spectra 
belonging to the various regions such that some spectra belonging to 
the east part 
also belong to the north (see Fig.~\ref{fig:nsew}). The EW is not 
meaningful to calculate for the different regions, as the continuum 
emission is restricted to the total emission. Line fluxes listed in 
column 3 have not been corrected for Galactic extinction.}
\label{tab:lya}
\end{table}

Previously, the existence of a disk in this system has been suggested from
studies of the metal absorption line profiles \citep{lu97}. This
interpretation is not supported by \citet{ledoux98}, who concluded that the
profiles could be caused by several interacting components, an interpretation
which is motivated by cold dark matter simulations of hierarchical clustering
of proto-galaxies \citep{haehnelt98}. Under the assumption of a rotating disk,
the enclosed mass calculated from \(M_{\mathrm{dyn}}=v_c^2 r / G\) is
$\sim10^{10}$ M$_{\odot}$. As will be discussed below, the \lya\ emission is
unlikely to originate in a rotating disk, and the mass calculated here is
probably overestimated.

\subsection{The double peaked \lya\ line}
\label{double}
\begin{figure}
  \hspace*{-0.7cm}\resizebox{9.7cm}{!}{\includegraphics[angle=90]{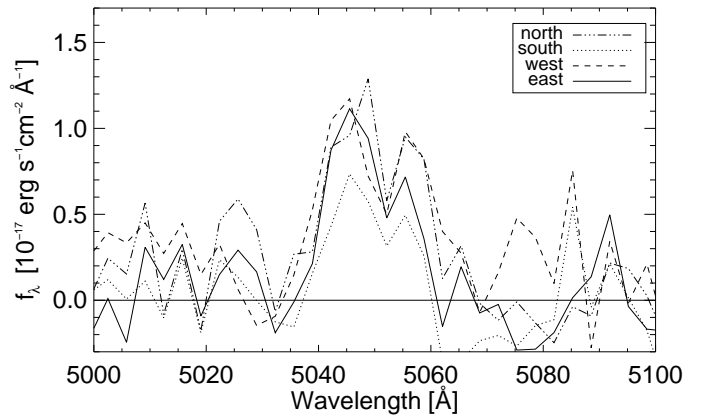}}
\caption{Spectra of the four different regions of the emission line
  object. The north, south, east, and west regions correspond to those
  listed in Table~\ref{tab:lya}. All regions are seen to have the same
  double peaked profile of the \lya\ line.}
\label{fig:sep}
\end{figure}

In Fig.~\ref{fig:spec} the \lya\ emission line appears to have a double peaked
profile. As this feature is not apparent in D96, we double checked carefully
for its integrity. Although the dip near the line centre is of the order of
the noise level, it appears in all spectra and therefore is almost certainly
real.  In Fig.~\ref{fig:sep} all the four separate spectra described above
have the same double peaked profile. The feature is not caused by the
subtraction of the sky background, nor is it due to imperfections in the CCD
in the region around the emission line.

The difference between our measured emission redshift and the one reported in
D96 may indicate that our wavelength calibration has an error of $\sim$1
{\AA}. This would not have an effect on the reported velocity difference since
the error would be systematic in all our spectra and not affect the
differences in the estimated Gaussian maxima. Furthermore, as mentioned, the
spectra are contaminated by the QSO, but a subtraction of a scaled QSO
spectrum from the four individual spectra does not change the \lya\ line
profile.

Anyhow, we caution the reader that other features appear to be systematic in
the spectra, e.g. a slightly smaller systematic depression is present in all
the spectra around 5020 {\AA} that could be due to an error in the background
estimation. Keeping this in mind, we note that a similar double peaked feature
was observed for the \lya\ line of the DLA galaxy in \object{Q2059--360}.
However, for this object the line profile changed with slit position
\citep{leibundgut99}.

The values for the centroids of the Gaussian fits listed in
Table~\ref{tab:lya} can thus be affected by unequal contributions from two
separate emission components.  We re-analysed the profiles by simultaneously
fitting two Gaussians resulting in centroids of the two peaks listed in column
1 and 2 in Table~\ref{tab:twopeaks}. For these two-Gaussian fits the widths of
the two individual components are barely resolved with this spectral
resolution.  The location of the first peak is shifted by 1.1~{\AA} from east
to west, corresponding to 65km~s$^{-1}$, while the second component is shifted
by --1{\AA}. Since the uncertainties for the estimate of the Gaussian
centroids are 0.5 {\AA}, the shifts are consistent with 0 within 1$\sigma$
errors.  The shifts for the north-south regions are likewise small
($<2.5${\AA} corresponding to $<150~$km~s$^{-1}$). For all regions the
splitting between the two Gaussians are 10--12.5 {\AA} corresponding to
velocities of 600--750~km~s$^{-1}$. Very likely, the velocity structure across
the extended \lya\ nebula is complex, and single Gaussian fits for such
extended areas as analysed here is over-simplistic.

Additionally, we have fit the profiles by a combination of a single Gaussian
emission line and a Gaussian absorption line creating the dip seen in the
spectra.  The centroids for the absorption line for these fits are listed in
Table~\ref{tab:twopeaks} in column 3.  The $\chi^2$s from these fits were
slightly smaller than for the two emission component fits, indicating that the
presence of an absorption line is preferred.

\begin{table}
\centering
\begin{tabular}{llll}
\\  \hline \hline  
\noalign{\smallskip}
region & centroid 1 & centroid 2  & abs. centroid \\
     &  {\AA}     & {\AA}    &    {\AA}\\
\noalign{\smallskip}
\hline
\noalign{\smallskip}
east    & 5045.9 &  5056.3 & 5051.1 \\
west    & 5044.8 &  5057.3 & 5051.0\\
south   & 5046.3 &  5056.3 & 5050.7\\
north   & 5046.5 &  5058.8 & 5052.2\\
\noalign{\smallskip}
\hline
\noalign{\smallskip}
\end{tabular}
\caption[]{The location of the two peaks from fitting simultaneously two
  Gaussians to the spectra in Fig.~\ref{fig:sep}. The centroids for the fits
  between the 4 regions vary with $<2.5$~{\AA} indicating small velocity 
  differences. For all regions the splitting between the two Gaussians 
  are 10--12 {\AA} corresponding to velocities of  600--750~km~s$^{-1}$. The
  third column lists the centroid for the absorption line for the fits
  consisting of an emission line and an absorption line.}
\label{tab:twopeaks}
\end{table}

\subsection{Artificial slit spectra}

 \begin{figure*}
\begin{minipage}[c]{.5\textwidth}
\resizebox{9cm}{!}{\includegraphics{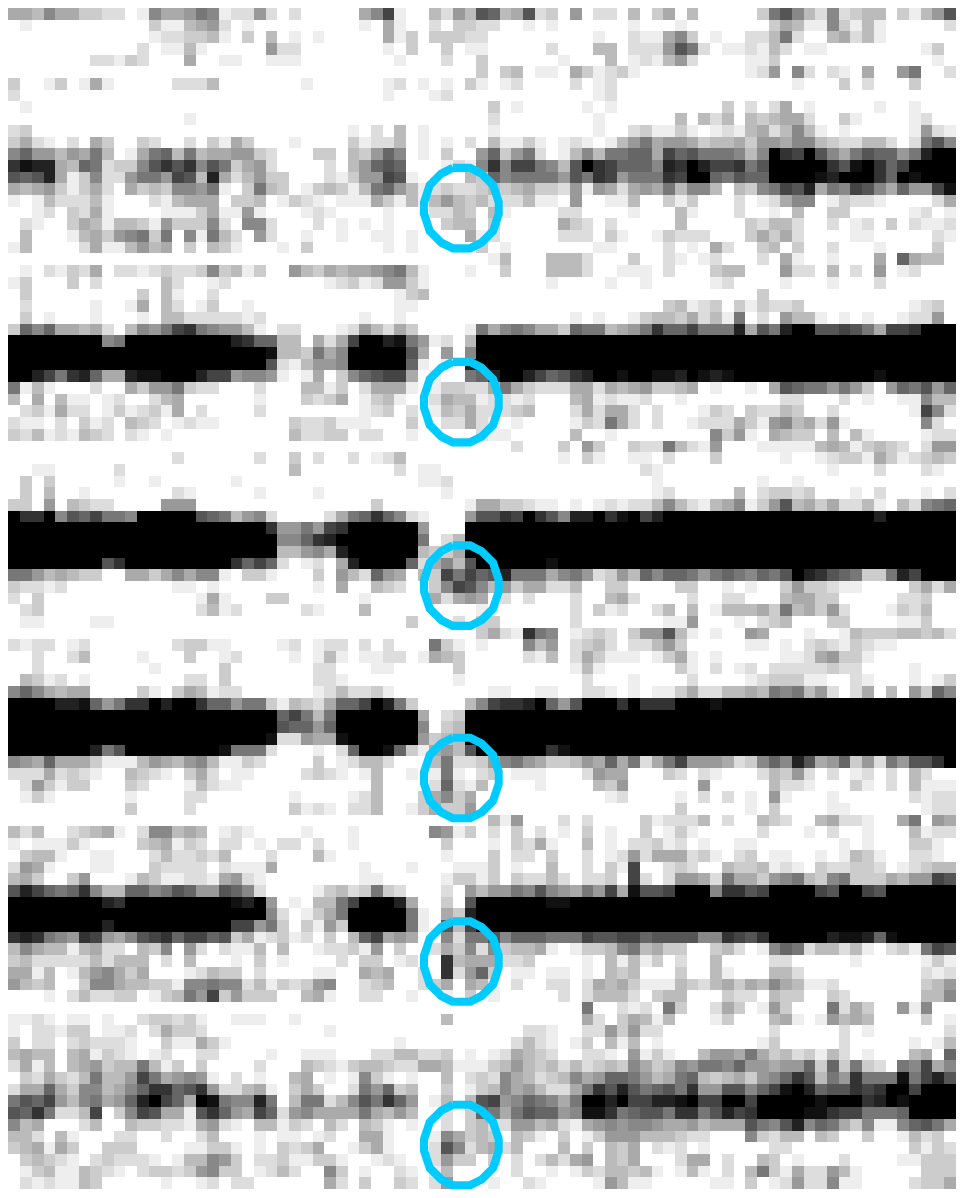}}
\end{minipage}%
\begin{minipage}[c]{.3\textwidth}
\resizebox{9cm}{!}{\includegraphics{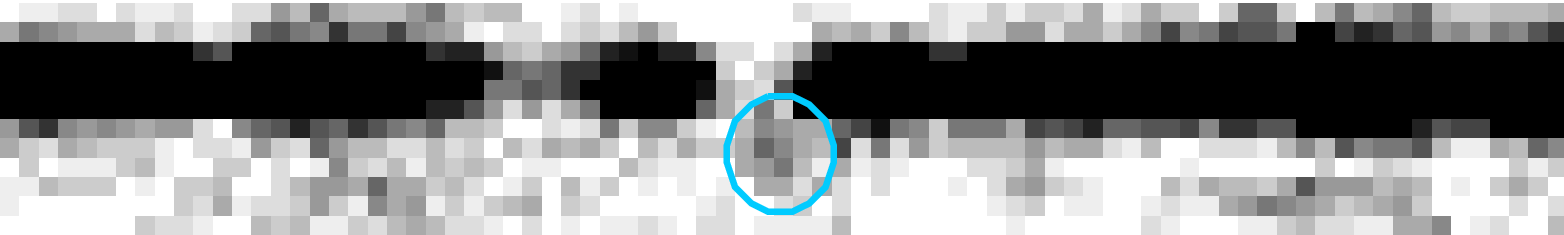}}
\resizebox{9cm}{!}{\includegraphics{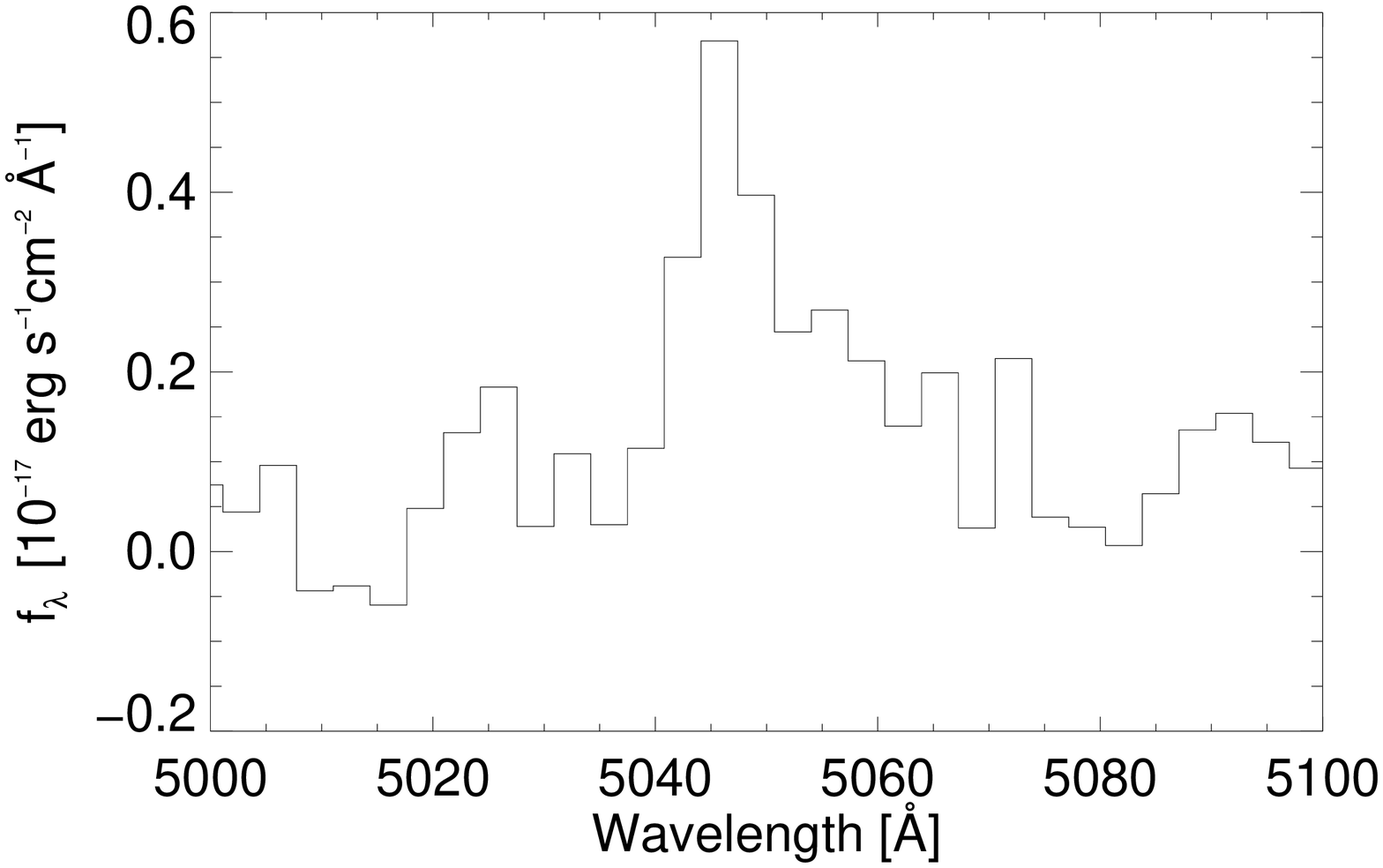}}
\end{minipage}
\caption{The left hand panel shows a section of the PMAS spectra consisting of
  spectra from 100 fibers around the region of the DLA. The spectra are
  oriented with a horizontal dispersion direction. Here are shown 6 groups of
  spectra where the emission from the extended nebula is visible within the
  small circles. Each group can be considered as representing a 0\farcs5 wide
  two-dimensional long slit spectrum. In comparison we have in the upper right
  hand panel shown an artificial long-slit spectrum one would have obtained
  with a 1\farcs0 slit placed in the north-south orientation as indicated in
  Fig~\ref{fig:wfpc}.  The lower right hand panel shows the corresponding
  one-dimensional spectrum from extracting the \lya\ emission spectrum. This
  is directly comparable to that in D96, in the sense that it exhibits the
  same red-winged profile, the same line flux, and the same line width. Note
  that the emission within the small circles do not overlap spatially with the
  QSO emission.}
\label{fig:art}
\end{figure*}

With integral field data we can reproduce the observations expected from slit
spectroscopy. In Fig~\ref{fig:art} we show a cut of the PMAS spectra around
the DLA line in the QSO, where we have indicated the emission from the
extended \lya\ nebula by small circles. In comparison we have in the upper
right hand panel created an artificial long slit spectrum with a width of
1\arcsec, and the corresponding one-dimensional spectrum is shown in the lower
panel. These plots can directly be compared to the ones presented in D96.  We
derive a line flux of $(6.5\pm1.0)\times10^{-17}$~\ecsa\ for the \lya\ line,
confirming the flux reported in D96, and a FWHM of $6.5\pm1.0$~{\AA},
corrected for the instrumental resolution, corresponding to a velocity of
$390\pm60$ km s$^{-1}$. This FWHM is similar to the the value in D96. With
these simple exercises we have shown that it is possible to reproduce the
previously published results derived from long-slit spectroscopy.

One sees furthermore in Fig.~\ref{fig:art}, that the \lya\ emission indicated
by the small circles do not overlap spatially with the location of the QSO.
This implies that the contours in Fig.~\ref{fig:map} at a distance less than
$1$\arcsec\ from the QSO only show emission from the QSO itself and not \lya\ 
emission from the DLA galaxy. Therefore we consider that selecting the
emission region from the DLA galaxy as indicated in Fig.~\ref{fig:nsew} is the
real extension of the \lya\ nebula associated with the DLA galaxy.

\section{HST deep imaging} 
\label{wfpc}
For comparison of the extended emission with high spatial resolution and
deeper optical data previously unpublished HST/WFPC2 images of Q2233+131 were
retrieved from the HST archive. Previously, a high spatial resolution image
from HST/STIS was presented by \citet{moller02} and we also retrieved the STIS
images from the archive for comparison. The WFPC2 planetary camera images were
obtained through the F702W filter with a total integration time of 10800~s and
the images were combined using the drizzle package in IRAF \citep{fruchter02}.
The resulting pixel scale in the drizzled image is 0\farcs023.

\begin{figure}
  \resizebox{\hsize}{!}{\includegraphics[]{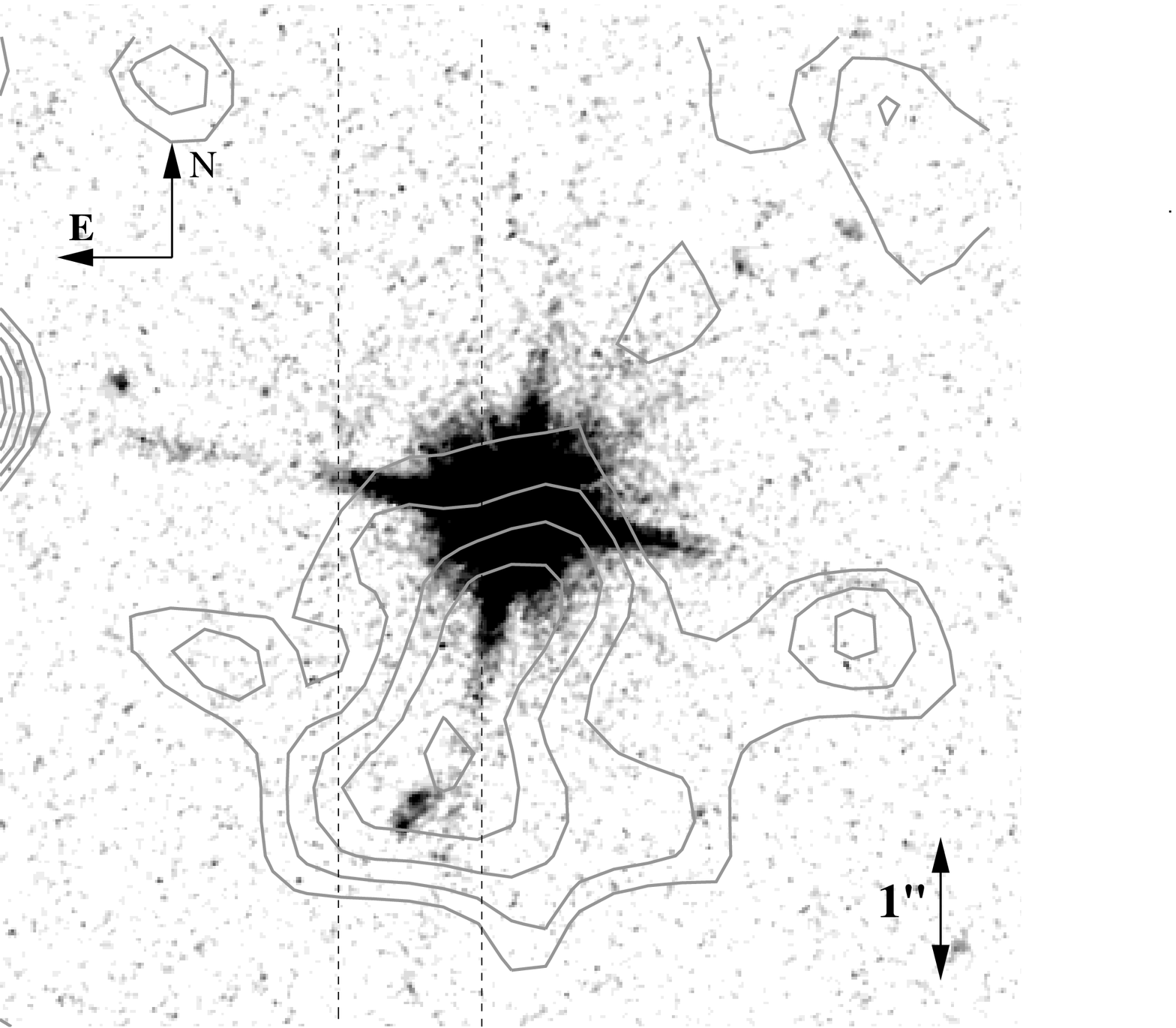}}
\caption{
  A negative 8\arcsec$\times$8\arcsec\ WFPC2 image of the Q2233+131.  North is
  up and east is left. The identified absorbing galaxy is seen towards the
  south of the QSO and the contours of the object in Fig.~\ref{fig:map} are
  overlaid. Note again, that the emission is a sum of the \lya\ emission from
  the nebula and some emission from the QSO. The levels of the contours
  correspond to 5--8$\sigma$ above the median background, with steps of
  0.75$\sigma$. There is a peak in the \lya\ emission close to the location of
  the DLA galaxy. Indicated by the vertical dashed lines are the position of
  the artificial slit from which the spectrum in Fig.~\ref{fig:art} was
  created}.
\label{fig:wfpc}
\end{figure}

An 8\arcsec$\times$8\arcsec\ section of the WFPC2 image is shown in
Fig.~\ref{fig:wfpc}.  The DLA galaxy is visible towards the south and the
contours of the extended \lya\ nebula detected by PMAS has been overplotted.
The exact positioning of the contours was checked from the knowledge of the
location of the QSO in the PMAS images from Fig.~\ref{fig:map}.  The \lya\ 
nebula is clearly extended, but it is difficult to measure the extension of it
in the direction towards the QSO since it is contaminated by the flux from the
red wing of the DLA line.  Avoiding this contamination requires an exact
knowledge of the point spread function behavior with wavelength such that the
flux contribution from the QSO can be subtracted. However as argued from
Fig.~\ref{fig:art}, there is no spatial overlap between the QSO and the \lya\ 
nebula so the contours to the north is caused by QSO emission only. In the
other direction towards the south there is a clear cutoff in the contours. We
therefore conclude that the \lya\ nebula is more elongated in the east--west
direction than in the north--south direction yielding the extension of the
\lya\ nebula of 3\arcsec$\times$5\arcsec. Considering that the observed
extension of the nebula is a convolution of the true emission and the seeing
during the observations, it could originate in smaller clouds in a more
complex environment.

In the WFPC2 image a faint and small galaxy is visible at the position
indicated in D96 and \citet{moller02}, and which is identical to the Lyman
Break Galaxy (LBG) detected by \citet{steidel95}. The impact parameter
measured in the image is 2\farcs37 corresponding to 18 kpc in the adopted
cosmology, the position angle is 159.1\deg, and the size of the galaxy along
the major axis is 0\farcs5, which at $z=3.15$ corresponds to 3.8~kpc.

In Fig.~\ref{fig:wfpc}, the size of the \lya\ nebula appears much larger than
the size of the DLA galaxy in the WFPC2 image. To quantify this we
investigated whether some low surface brightness emission is lost in the WFPC2
image due to different sensitivities. In the WFPC2 image the 3$\sigma$
limiting magnitude is 26.8~mag~arcsec$^{-2}$, corresponding to a flux limit of
$6\times10^{-17}$~\ecs~arcsec$^{-2}$, while the PMAS observations of the \lya\ 
object detect emission of 4$\times10^{-17}$\ecs~arcsec$^{-2}$, implying that
the detection limit of the PMAS spectra are roughly the same as for the WFPC2
images.  Therefore, if the continuum emission were as extended as the \lya\ 
line emission it would have been detected in the HST images.

The optical magnitude of the LBG corresponds to an $L^*$ galaxy (D96). In
Fig.~\ref{fig:wfpc} the DLA galaxy is seen to be composed of two components
separated by 0\farcs25 corresponding to $\sim$2 kpc at the redshift of the DLA
galaxy, and the two components are aligned almost orthogonal to the long axis
of the extended emission. A galactic outflow will naturally occur along the
minor axis of the galaxy as described in \citet{heckman90}, which would
explain the orientation of the elongated \lya\ nebula with respect to the LBG
orientation. The irregular morphology of the galaxy was also found by
\cite{moller02} in their HST/STIS data.  They noted that the morphology of the
DLA galaxy was not unusual compared to field galaxies at the same redshift.

Using aperture photometry in IRAF we find the magnitude of the DLA galaxy
$m_{\mathrm{F702W}}~=~24.80\pm0.1$ using the planetary camera zero point from
the WFPC2 Instrument Handbook. This corresponds to a continuum flux of
$\sim$3$\times10^{-19}$~\ecsa\ at $\lambda~=~1690$~{\AA} in the restframe of the
DLA galaxy. This is below our detection limit in the PMAS spectra and explains
the non-detection of underlying continuum emission from the DLA galaxy. Dust
obscuration would make the intrinsic luminosity only larger. The derived flux
from the WFPC2 image, confirming that the UV continuum is consistent with
being flat, is in agreement with the hypothesis in D96 that the galaxy is very
young, and in a star-bursting phase. Using the same aperture for the STIS
data we derive $m(V_{50})=25.65\pm0.1$ in agreement within 1$\sigma$ of the
value reported in \citet{moller02}.

\section{\bf{\lya}\ emission from high redshift objects}
\label{hi_z}
We now compare the properties of other DLA galaxies with those for the
Q2233+131 DLA galaxy.  

The EW of the observed \lya\ line is relatively large compared to that of
other high redshift galaxies \citep{shapley03}.  However, one must note that
using a 1\arcsec\ long slit spectrum one would have obtained a 4 times smaller
flux, and therefore decreased the inferred EW by the same amount.  Values of
\lya\ EWs for high redshift \lya\ emitters are $>14$ {\AA}, while some may
have EWs which are two orders of magnitude larger \citep{kudritzki00}. For the
\lya\ emission from the DLA galaxies in \object{PKS~0528--250} and
\object{Q2206--1958} the measured EWs are 63~{\AA} and 83~{\AA}, respectively,
which is typical for LBGs at the same redshifts \citep{moller02}.  

We found that the line width for the Q2233+131 DLA galaxy suggest larger
velocities: 1000~km~s$^{-1}$ compared to $\sim$700~km~s$^{-1}$ observed for
other DLA galaxies. An explanation for this may be that we have co-added the
spectra over a large area. Velocity differences of 300~km~s$^{-1}$ have been
found from one end to the other of the extended emission, and with the
artificial slit spectra created from the integral field spectra we indeed find
a smaller value.

\citet{leibundgut99} found evidence for an extended \lya\ emission nebula
associated with the DLA in \object{Q2059--360} using long-slit spectroscopy,
with the slit placed at several positions offset from the QSO.  They found a
velocity difference between the \lya\ emission line and the DLA absorption of
+490~km~s$^{-1}$, and that the \lya\ emission line could be described by two
components separated by 5~{\AA}. However, they could not exclude that the
\lya\ emission was affected by the QSO, which has a very small redshift
difference from the \lya\ emission line. Nevertheless, these observations are
remarkably similar to those presented here of the Q2233+131 DLA, except for
the fact that the QSO in our case is well separated in redshift space by
$\Delta z=0.15$.

The properties derived for the Q2233+131 \lya\ nebula are also similar to the
DLA galaxy detected in \object{Q0151+048A} which is described in
\citet{fynbo99} and \citet{fynbo00}. They concluded that some of the extended
\lya\ emission could be caused by photoionization by the QSO, which has the
same redshift as the DLA. This is not the case here.  The proper distance
between the QSO and the DLA is 120~Mpc in the adopted cosmology. Using the
relation between the QSO absolute magnitude and the distance to the DLA given
in \citet{warren96}, we find that the ionizing flux can be at most
$10^{-21}$~\ecs~arcsec$^{-2}$, which is 4 orders of magnitude below the
observed value.

\citet{warren96} found evidence that the DLA galaxy termed S1 in
\object{Q0528--250} has an extension of 1\arcsec\ after correcting for
atmospheric seeing suggesting that \lya\ emission is more extended than the
region emitting continuum radiation.

Previously a couple of LBGs at $z=3.1$ have been associated with surrounding
\lya\ emission nebulae in \citet{steidel00}, who termed these ``\lya\ blobs''.
The properties of the extended nebula presented here are less extreme, but not
very different from the \lya\ blobs.  The total luminosity of the \lya\ 
emission from the DLA nebula is a factor 10 less than from the \lya\ blobs, and
the line width of the DLA emission is a factor of 2 smaller \citep{ohyama03}.
The inferred size of the DLA emission is smaller than for the \lya\ blobs,
which have sizes of $\sim15$\arcsec, i.e.  a factor of 3 larger than the DLA's
emission, but much of the emission is diffuse, and substructure in the blobs is
clearly visible.

It is likely that the observed extension of the DLA galaxy's \lya\ emission
becomes larger as one goes to fainter fluxes, and since the surface brightness
decreases with redshift as $(1+z)^4$, it is difficult to detect faint extended
objects especially with a 4m class telescope.


\section{Origin of the extended emission}
\label{origin}
We address now the question whether the \lya\ emission is induced by a
superwind from the DLA galaxy and caused by star formation, or if it
originates in a rotating disk.

For the assumption of a rotating disk the relation \(v_c~=~v(\mathrm{FWHM})/(2
\sin i)\) gives \( v_c~=~(545\pm65)/\sin i\).  Such a large value makes the
interpretation of a rotating disk questionable. A more likely explanation is
resonance scattering of \lya\ photons that can also produce a large FWHM of
the emission line. The resonant nature of the \lya\ photon neither increases
nor decreases the total flux emitted by the source, only the escape direction,
but when the dust-to-gas ratio of the environment exceeds 10\% of the Galactic
value, the extinction will be significant due to the increased escape path
from the resonance scattering of the \lya\ photons.  While in dust free
star-forming galaxies the expected \lya\ EW is 100--200 {\AA}
\citep{charlot93}, the fact that the measured EW is similar to this value
suggests that the most probable source of ionization is massive star
formation, and that dust extinction plays a small role.

The red wing of the line profile observed by D96 is marginally detected in our
lower resolution data.  We can reproduce the line profile in the artificial
slit spectrum in Fig.~\ref{fig:art}, but it is much weaker when co-adding
several spaxels as in Fig.~\ref{fig:spec} indicating the presence of some
velocity structure.

\subsection{\lya\ emission indicative of star formation}

In star-forming galaxies, it is mainly the OB stars that produce Lyman
continuum photons. These photons will be absorbed by the surrounding neutral
hydrogen and re-emitted as line photons.  \citet{kennicutt98} gives the
relation between the H$\alpha$ luminosity ($L$) of a galaxy and its
star-formation rate (SFR). Assuming case B recombination \citep{osterbrock89}
and that the expected ratio of \lya\ to H$\alpha$ flux is
\(F(\lya)/F(\textrm{H}\alpha)\approx 10 \), implies that
\[\textrm{SFR~(M}_{\odot}~\textrm{yr}^{-1})=L(\lya)/1.26\times10^{42}~\textrm{erg~s}^{-1}\]
The conversion factor depends strongly on the adopted initial mass function,
and we adopt an uncertainty of 50\%. The relation between the SFR and the
luminosity also depends on the escape fraction of the \lya\ photons, which is
very uncertain as the extinction is unknown.

Assuming that the \lya\ emission is caused by star formation we find
$\textrm{SFR}=19\pm10$~M$_{\odot}$~yr$^{-1}$, not corrected for the unknown
escape fraction. This is a typical value compared to the SFRs found for LBGs
having strong \lya\ emission \citep{shapley03}.  It is also consistent with
the observed continuum, as can be shown using the conversion from the
continuum flux at 1500{\AA} in the restframe of a galaxy and the SFR, for
which \citet{madau98} gives the relation
SFR(M$_{\odot}~\textrm{yr}^{-1})=1.3\times10^{-28} L_{\nu}$
(erg~s$^{-1}$~Hz$^{-1}$). We take into account an uncertainty for the
conversion factor of $\sim$30\%.  The observed $R$ band corresponds to
1500~{\AA} in the rest frame of the DLA system.  Correcting the $R$ band
magnitude for Galactic extinction and applying an offset converting Vega
magnitudes to AB magnitudes \citep{fukugita95} cancels each other out, i.e.
$R_{AB}=24.8\pm1$.  Using the relation above this magnitude corresponds to a
SFR$=12\pm5$~M$_{\odot}~\textrm{yr}^{-1}$, which is consistent with the SFR
derived from the \lya\ flux.

Given the uncertainties regarding the conversions from fluxes to SFRs, we find
that the \lya\ emission around the DLA galaxy could be caused by massive
stars.  In principle a significant amount of dust would imply that the true
\lya\ flux is higher than derived from the UV flux. On the other hand, the
consistency between the two measurements indicates that dust extinction plays
no major role in this particular case.  A low dust content could imply a
larger escape fraction of the \lya\ photons.  Another explanation of the large
\lya\ luminosity could be the complex dynamics involved in this system
\citep{dawson02}, and not directly related to the dust content as argued for
nearby galaxies \citep{kunth03}.

\subsection{Origin of the double peaked emission line}
\label{d2}
Several authors have analysed the transfer of \lya\ photons in neutral clouds
\citep{adams72,urbaniak81,zheng02}. Even in the case of a static cloud, the
emission profile of \lya\ would have a double peak, whereas an outflow would
blend the two components of the line. The wavelength difference between the
two components is \(\Delta\lambda~=~\sqrt{\frac{2kT}{m_{H} c^2}}\lambda_0\),
where $k$ is the Boltzmann constant, $T$ is the temperature, $m_H$ is the
Hydrogen mass, $c$ is the speed of light, and $\lambda_0$ is the rest frame
wavelength of \lya.  Assuming a temperature of $10^4$~K, the corresponding
splitting of the line is 0.2 {\AA} at $z=3.15$, which is much smaller than the
observed splitting. On the other hand, given the observed splitting, a
temperature of $3\times10^7$ K is required, which is two orders of magnitude
higher than temperatures in superwind outflows at distances far away from the
central starburst \citep{heckman90}.  We therefore argue that the observed
double peak is not caused by a static neutral cloud, but has a more complex
origin.

In the case of an outflow, the double peaked emission line can be explained.
A galactic wind, powered by star formation and supernova explosions, expands
outwards and interacts with the surrounding material, shock heats and
accelerates this, producing emission line filaments. The preferred direction
of expansion is where the pressure gradient is largest, i.e. along the
rotation axis of the galaxy usually aligned with the minor axis.  The wind
creates a large shell of swept up material, and observations towards such an
object would intersect the shell in the front and the rear end giving rise to
a double peaked emission line profile. Such double peaked profiles of optical
emission lines have been observed for nearby starburst galaxies which exhibit
these large scale outflows typically termed superwinds \citep{heckman90}.
Additionally, a double peaked \lya\ emission line profile was found from the
starburst galaxy \object{T1214--277} by \citet{mashesse03}, who concluded that
the feature is caused by emission in an outflow.  The inferred velocity
separation between the two peaks from the DLA galaxy emission are similar to
the highest ones measured for the nearby galaxies in \citet{heckman90} and
similar to the \lya\ emission line splitting from T1214--277
\citep{mashesse03}.

\subsection{The emitter and the absorber}
In the case of a superwind outflow, the DLA cloud is affected by its impact,
or could event be created by a such. \citet{taniguchi01} found that a
superwind can create Lyman limit systems with $N_H~>~10^{19}$~cm$^{-2}$ along
filaments. Followed by gravitational collapse along such filaments DLA clouds
can be created.  Several authors have pointed out that there could be a
relation between superwinds from galaxies and the metal absorption systems
seen in quasar spectra \citep[e.g.][]{heckman90,nulsen98}, thus explaining the
early chemical enrichment of neutral gas clouds seen at the highest redshifts.
However, \citet{bond01} conclude the opposite, on the basis of \ion{Mg}{II}
absorption profiles, that superwinds are not causing the majority of DLA
systems.

The velocity difference between the emitter and the absorber could be caused
by the impact of a superwind from the galaxy which would accelerate a
pre-existing low temperature clouds in the surrounding environment. Depending
on the initial distance from the starburst, column density of the neutral gas
cloud, and the luminosity of the starburst, it can be shown that the expected
acceleration is of the order of a few times 100~km~s$^{-1}$ \citep{heckman90}.
This would be consistent with the velocity difference of $\sim$300~km~s$^{-1}$
between the \lya\ emission and the DLA absorption measured here.

\section{Mass estimate}
\label{mass}
When the \lya\ luminosity is known, the mass contained in the extended nebula
can be evaluated following the method described in \citet{morse98}. The
luminosity is given by
\begin{equation}
L_{\lya} = j_{\lya} V f \quad \textrm{[erg s}^{-1}]
\end{equation}
where $j_{\lya}$ is the volume emissivity, $V$ is the volume, and $f$ is the
volume filling factor. For gas with a temperature of $T=10^4$ K assuming
case B recombination, \citet{osterbrock89} gives
\begin{equation}
j_{\lya} = 3.56 \times 10^{-24} n_p n_e  \quad \textrm{[erg cm}^{-3} \textrm{s}^{-1}]
\label{eq:ne}
\end{equation}
where $n_p$ and $n_e$ are the number densities of protons and electrons
respectively. The total mass of ions in the nebula assuming negligible metal
content is
\begin{equation}
M_{\mathrm{ion}} = (n_p m_p + n_{He} m_{He}) V f 
\end{equation}
Using the approximations \(n_p=0.9 n_e\) and \(n_{He}=0.1 n_p\), this reduces
to
\begin{eqnarray}
M_{\mathrm{ion}}& =& 1.26 n_e m_p V f \nonumber \\
       & \approx& 1.06 \times 10^{-57} n_e V f  \quad [M_{\odot}]  
\label{eq:ion}
\end{eqnarray}
Combining the Equations~\ref{eq:ne} and \ref{eq:ion} gives
\begin{equation}
M_{\mathrm{ion}} \approx 5.91 \times 10^{-46} (L_{\lya} V f)^{1/2}
\label{eq:mass}
\end{equation}
For calculating the volume of the \lya\ nebula, we make an approximation by a
cylinder with a radius of 10 kpc and a height of 40 kpc. With the calculated
luminosity given above we find
$M_{\mathrm{ion}}=~5.7~\times~10^{10}~(f)^{1/2}~M_{\odot}$. Assuming
$f=10^{-5}$ taken as a rough guess, along with the arguments in
\citet{mccarthy90}, we find $M_{\mathrm{ion}}=1.8\times10^{8} M_{\odot}$.

Due to the unknown dust content in the cloud the unextincted \lya\ luminosity
could be much higher than derived, and consequently also the mass. On the
other hand, a smaller filling factor would decrease the derived mass, i.e.
these two unknown factors may even compensate each other to some extent.


\section{Conclusions}
\label{conc}
Using integral field spectra obtained with PMAS, we have presented evidence
that the DLA galaxy previously known to be responsible for the DLA system in
Q2233+131 has an extended envelope of \lya\ emission. From the spectra we
constructed an artificial narrow-band image which showed that the extension of
the emission is 3\arcsec$\times$5\arcsec\ corresponding to 23$\times$38 kpc.
This cloud has a line flux of $(2.8\pm0.3)\times10^{-16}$ \ecs\ corrected for
Galactic extinction, corresponding to a luminosity of
$(2.4_{-0.2}^{+0.3})\times10^{43}$~erg~s$^{-1}$ at $z=3.1538$. We derive the
rest frame EW$ =190_{-70}^{+150}$ {\AA}.
 
The source of the ionization is likely star formation within the DLA galaxy.
The measurement of the \lya\ luminosity is generally considered as a very
uncertain method for estimating the SFR given the unknown dust obscuration and
escape fractions. With this in mind we find a
$\textrm{SFR}~=~19\pm10$~M$_{\odot}$~yr$^{-1}$ from the \lya\ luminosity, a
result which is consistent with the
$\textrm{SFR}~=~12\pm5$~M$_{\odot}$~yr$^{-1}$ derived from the UV continuum
flux from the observed ground based $R$ band magnitude.  The agreement between
the two suggests that dust extinction plays only a small role.

A velocity offset of $\sim$270$\pm$40 km s$^{-1}$ between the emission and the
absorption component in the DLA galaxy does not support the hypothesis that
the absorbing cloud resides in a rotating disk. We investigated the velocity
structure of the extended \lya\ emission further by constructing 4 composite
spectra.  By fitting Gaussian profiles to spectra from the eastern and
western region we found an offset of 2.5~{\AA}, which corresponds to a
velocity difference of 150~km~s$^{-1}$.  The splitting of the spectra into a
northern and a southern region gave a velocity difference of 300~km~s$^{-1}$.
These results combined with the extended nature of the \lya\ nebula is not
consistent with the properties of a disk similar to that in large spiral
galaxies.

The object's characteristics (luminosity, FWHM, and spatial extension) are
similar to that for some LBGs for which extended \lya\ emission has been
detected. These high redshift \lya\ emitters are thought to be created by an
outflowing wind. Recombination of ionised hydrogen creates \lya\ photons that
are resonantly scattered in the environment, which results in the observed
extended envelope.  We therefore consider the interpretation of an outflow
from the galaxy a more likely interpretation. Furthermore, a galactic outflow
can create emission lines with double peaked profiles in agreement with the
observations.

We argued that the DLA cloud is not a part of a rotating disk surrounding the
DLA galaxy. The relation between the extended \lya\ emission and the DLA cloud
in the line of sight towards the QSO is somewhat speculative. The wind
responsible for the extended emission can create neutral clouds with high
column densities located along filamentary structures. Gravitational collapse
in these filaments can create DLA clouds. Another explanation could be a
previously existing neutral gas cloud which has been accelerated by the
superwind giving rise to a velocity separation of 300 km s$^{-1}$. A third
explanation is that the DLA cloud could be an otherwise unrelated infalling
cloud. This third explanation is questionable, given the fact that the
measured metallicity in the DLA cloud suggests some processing of stellar
material.

We have demonstrated the advantages of using integral field spectroscopy for
investigating DLA systems in the terms of confirming the galaxies responsible
for the DLA absorption. First of all, one can avoid the conventional two step
procedure for detecting candidate DLA galaxies near the line of sight towards
the QSOs, which later have to be followed up spectroscopically.  With integral
field spectra one can create artificial narrow-band images suited to any
wavelengths required.  Secondly, a non-extended line emitting region could be
missed by placing the slit in a less favorable angle, while in the case of
an extended emission the line fluxes would inevitably be underestimated.

We will undertake further investigation of the Q2233+131 DLA system with PMAS
in order to investigate the velocity structure of the \lya\ nebula.
Specifically, the system should be observed with a higher spectral resolution
which should allow to determine the presence of P Cygni profiles expected for
the case of superwind outflows but also in order to investigate the double
peaked profile in detail.

\begin{acknowledgements}
  L.~Christensen acknow\-ledges support by the German Verbundforschung
  associated with the ULTROS project, grant no. 05AE2BAA/4. S.F.~S\'anchez
  acknowledges the support from the Euro3D Research Training Network, grant
  no. HPRN-CT2002-00305. K. Jahnke and L. Wisotzki acknowledge a DFG travel
  grant under Wi~1369/12-1.  L. \v C. Popovi\'c acknowledges support by
  Alexander von Humboldt Foundation through the program for foreign scholars
  and the Ministry of Science, Technologies and Development of Serbia through
  the project "Astrophysical Spectroscopy of Extragalactic Objects". We thank
  Palle M{\o}ller for his comments and suggestions on an earlier version of
  this paper.
\end{acknowledgements}

\bibliography{DLA2233}

\begin{thebibliography}{55}
\expandafter\ifx\csname natexlab\endcsname\relax\def\natexlab#1{#1}\fi

\bibitem[{{Adams}(1972)}]{adams72}
{Adams}, T.~F. 1972, \apj, 174, 439

\bibitem[{{Bechtold}(1994)}]{bechtold94}
{Bechtold}, J. 1994, \apjs, 91, 1

\bibitem[{Becker(2002)}]{becker01}
Becker, T. 2002, PhD thesis, Astrophysikalisches Institut Potsdam, Germany

\bibitem[{{Bond} {et~al.}(2001){Bond}, {Churchill}, {Charlton}, \&
  {Vogt}}]{bond01}
{Bond}, N.~A., {Churchill}, C.~W., {Charlton}, J.~C., \& {Vogt}, S.~S. 2001,
  \apj, 562, 641

\bibitem[{{Charlot} \& {Fall}(1993)}]{charlot93}
{Charlot}, S. \& {Fall}, S.~M. 1993, \apj, 415, 580

\bibitem[{{Colbert} \& {Malkan}(2002)}]{colbert02}
{Colbert}, J.~W. \& {Malkan}, M.~A. 2002, \apj, 566, 51

\bibitem[{{Curran} {et~al.}(2002){Curran}, {Webb}, {Murphy}, {Bandiera},
  {Corbelli}, \& {Flambaum}}]{curran02}
{Curran}, S.~J., {Webb}, J.~K., {Murphy}, M.~T., {et~al.} 2002, Publications of
  the Astronomical Society of Australia, 19, 455

\bibitem[{{Dawson} {et~al.}(2002){Dawson}, {Spinrad}, {Stern}, {Dey}, {van
  Breugel}, {de Vries}, \& {Reuland}}]{dawson02}
{Dawson}, S., {Spinrad}, H., {Stern}, D., {et~al.} 2002, \apj, 570, 92

\bibitem[{{Djorgovski} {et~al.}(1996){Djorgovski}, {Pahre}, {Bechtold}, \&
  {Elston}}]{djorgovski96}
{Djorgovski}, S.~G., {Pahre}, M.~A., {Bechtold}, J., \& {Elston}, R. 1996,
  \nat, 382, 234

\bibitem[{{Ellison} {et~al.}(2002){Ellison}, {Yan}, {Hook}, {Pettini}, {Wall},
  \& {Shaver}}]{ellison02}
{Ellison}, S.~L., {Yan}, L., {Hook}, I.~M., {et~al.} 2002, \aap, 383, 91

\bibitem[{{Filippenko}(1982)}]{filippenko02}
{Filippenko}, A.~V. 1982, PASP, 94, 715

\bibitem[{{Fruchter} \& {Hook}(2002)}]{fruchter02}
{Fruchter}, A.~S. \& {Hook}, R.~N. 2002, PASP, 114, 144

\bibitem[{{Fukugita} {et~al.}(1995){Fukugita}, {Shimasaku}, \&
  {Ichikawa}}]{fukugita95}
{Fukugita}, M., {Shimasaku}, K., \& {Ichikawa}, T. 1995, \pasp, 107, 945

\bibitem[{{Fynbo} {et~al.}(2000){Fynbo}, {Burud}, \& {M{\o}ller}}]{fynbo00}
{Fynbo}, J.~U., {Burud}, I., \& {M{\o}ller}, P. 2000, \aap, 358, 88

\bibitem[{{Fynbo} {et~al.}(1999){Fynbo}, {M{\o}ller}, \& {Warren}}]{fynbo99}
{Fynbo}, J.~U., {M{\o}ller}, P., \& {Warren}, S.~J. 1999, \mnras, 305, 849

\bibitem[{{Haehnelt} {et~al.}(1998){Haehnelt}, {Steinmetz}, \&
  {Rauch}}]{haehnelt98}
{Haehnelt}, M.~G., {Steinmetz}, M., \& {Rauch}, M. 1998, \apj, 495, 647

\bibitem[{{Heckman} {et~al.}(1990){Heckman}, {Armus}, \& {Miley}}]{heckman90}
{Heckman}, T.~M., {Armus}, L., \& {Miley}, G.~K. 1990, \apjs, 74, 833

\bibitem[{{Hopp} \& {Fernandez}(2002)}]{hopp02}
{Hopp}, U. \& {Fernandez}, M. 2002, Calar Alto Newsletter No.4,
  http://www.caha.es/newsletter/news02a/hopp/paper.pdf

\bibitem[{{Hunstead} {et~al.}(1990){Hunstead}, {Fletcher}, \&
  {Pettini}}]{hunstead90}
{Hunstead}, R.~W., {Fletcher}, A.~B., \& {Pettini}, M. 1990, \apj, 356, 23

\bibitem[{{Kennicutt}(1998)}]{kennicutt98}
{Kennicutt}, R.~C. 1998, ARA\&A, 36, 189

\bibitem[{{Kudritzki} {et~al.}(2000){Kudritzki}, {M{\' e}ndez}, {Feldmeier},
  {Ciardullo}, {Jacoby}, {Freeman}, {Arnaboldi}, {Capaccioli}, {Gerhard}, \&
  {Ford}}]{kudritzki00}
{Kudritzki}, R.-P., {M{\' e}ndez}, R.~H., {Feldmeier}, J.~J., {et~al.} 2000,
  \apj, 536, 19

\bibitem[{{Kunth} {et~al.}(2003){Kunth}, {Leitherer}, {Mas-Hesse}, {\"Ostlin},
  \& {Petrosian}}]{kunth03}
{Kunth}, D., {Leitherer}, C., {Mas-Hesse}, M., {\"Ostlin}, G., \& {Petrosian},
  A. 2003, ApJL submitted (astro-ph/0307555)

\bibitem[{{Le Brun} {et~al.}(1997){Le Brun}, {Bergeron}, {Boisse}, \&
  {Deharveng}}]{lebrun97}
{Le Brun}, V., {Bergeron}, J., {Boisse}, P., \& {Deharveng}, J.~M. 1997, \aap,
  321, 733

\bibitem[{{Ledoux} {et~al.}(1998){Ledoux}, {Petitjean}, {Bergeron}, {Wampler},
  \& {Srianand}}]{ledoux98}
{Ledoux}, C., {Petitjean}, P., {Bergeron}, J., {Wampler}, E.~J., \& {Srianand},
  R. 1998, \aap, 337, 51

\bibitem[{{Leibundgut} \& {Robertson}(1999)}]{leibundgut99}
{Leibundgut}, B. \& {Robertson}, J.~G. 1999, \mnras, 303, 711

\bibitem[{{Lu} {et~al.}(1997){Lu}, {Sargent}, \& {Barlow}}]{lu97}
{Lu}, L., {Sargent}, W.~L.~W., \& {Barlow}, T.~A. 1997, \apj, 484, 131

\bibitem[{{Madau} {et~al.}(1998){Madau}, {Pozzetti}, \& {Dickinson}}]{madau98}
{Madau}, P., {Pozzetti}, L., \& {Dickinson}, M. 1998, ApJ, 498, 106

\bibitem[{{Mas-Hesse} {et~al.}(2003){Mas-Hesse}, {Kunth}, {Tenario-Tagle},
  {Leitherer}, {Terlevich}, \& {Terlevich}}]{mashesse03}
{Mas-Hesse}, J.~M., {Kunth}, D., {Tenario-Tagle}, G., {et~al.} 2003,
  (astro-ph/0309396)

\bibitem[{{McCarthy} {et~al.}(1990){McCarthy}, {Spinrad}, {Dickinson}, {van
  Breugel}, {Liebert}, {Djorgovski}, \& {Eisenhardt}}]{mccarthy90}
{McCarthy}, P.~J., {Spinrad}, H., {Dickinson}, M., {et~al.} 1990, \apj, 365,
  487

\bibitem[{{M{\o}ller} \& {Warren}(1993)}]{moller93}
{M{\o}ller}, P. \& {Warren}, S.~J. 1993, \aap, 270, 43

\bibitem[{{M{\o}ller} {et~al.}(2002){M{\o}ller}, {Warren}, {Fall}, {Fynbo}, \&
  {Jakobsen}}]{moller02}
{M{\o}ller}, P., {Warren}, S.~J., {Fall}, S.~M., {Fynbo}, J.~U., \& {Jakobsen},
  P. 2002, \apj, 574, 51

\bibitem[{{Morse} {et~al.}(1998){Morse}, {Cecil}, {Wilson}, \&
  {Tsvetanov}}]{morse98}
{Morse}, J.~A., {Cecil}, G., {Wilson}, A.~S., \& {Tsvetanov}, Z.~I. 1998, \apj,
  505, 159

\bibitem[{{Nulsen} {et~al.}(1998){Nulsen}, {Barcons}, \& {Fabian}}]{nulsen98}
{Nulsen}, P.~E.~J., {Barcons}, X., \& {Fabian}, A.~C. 1998, \mnras, 301, 168

\bibitem[{{Ohyama} {et~al.}(2003){Ohyama}, {Taniguchi}, {Kawabata}, {Shioya},
  {Murayama}, {Nagao}, {Takata}, {Iye}, \& {Yoshida}}]{ohyama03}
{Ohyama}, Y., {Taniguchi}, Y., {Kawabata}, K.~S., {et~al.} 2003, \apjl, 591, L9

\bibitem[{{Osterbrock}(1989)}]{osterbrock89}
{Osterbrock}, D.~E. 1989, {Astrophysics of gaseous nebulae and active galactic
  nuclei} (Mill Valley, CA, University Science Books)

\bibitem[{{Petitjean} {et~al.}(1996){Petitjean}, {Pecontal}, {Valls-Gabaud}, \&
  {Charlot}}]{petitjean96}
{Petitjean}, P., {Pecontal}, E., {Valls-Gabaud}, D., \& {Charlot}, S. 1996,
  \nat, 380, 411

\bibitem[{{Prochaska} {et~al.}(2003){Prochaska}, {Gawiser}, {Wolfe}, {Castro},
  \& {Djorgovski}}]{prochaska03}
{Prochaska}, J., {Gawiser}, E., {Wolfe}, A., {Castro}, S., \& {Djorgovski},
  S.~G. 2003, ApJL, 595, L9

\bibitem[{{Rao} \& {Turnshek}(2000)}]{rao00}
{Rao}, S.~M. \& {Turnshek}, D.~A. 2000, \apjs, 130, 1

\bibitem[{{Roth} {et~al.}(2000){Roth}, {Bauer}, {Dionies}, {Fechner}, {Hahn},
  {Kelz}, {Paschke}, {Popow}, {Schmoll}, {Wolter}, {Laux}, \&
  {Altmann}}]{pmas00}
{Roth}, M.~M., {Bauer}, S., {Dionies}, F., {et~al.} 2000, in Proc. SPIE, Vol.
  4008, 277--288

\bibitem[{{S\'anchez}(2003)}]{sanchez03}
{S\'anchez}, S.~F. 2003, AN accepted (astro-ph/0310677)

\bibitem[{{Sargent} {et~al.}(1989){Sargent}, {Steidel}, \&
  {Boksenberg}}]{sargent89}
{Sargent}, W.~L.~W., {Steidel}, C.~C., \& {Boksenberg}, A. 1989, \apjs, 69, 703

\bibitem[{{Schlegel} {et~al.}(1998){Schlegel}, {Finkbeiner}, \&
  {Davis}}]{schlegel98}
{Schlegel}, D.~J., {Finkbeiner}, D.~P., \& {Davis}, M. 1998, \apj, 500, 525

\bibitem[{{Shapley} {et~al.}(2003){Shapley}, {Steidel}, {Pettini}, \&
  {Adelberger}}]{shapley03}
{Shapley}, A.~E., {Steidel}, C.~C., {Pettini}, M., \& {Adelberger}, K.~L. 2003,
  \apj, 588, 65

\bibitem[{{Steidel} {et~al.}(2000){Steidel}, {Adelberger}, {Shapley},
  {Pettini}, {Dickinson}, \& {Giavalisco}}]{steidel00}
{Steidel}, C.~C., {Adelberger}, K.~L., {Shapley}, A.~E., {et~al.} 2000, \apj,
  532, 170

\bibitem[{{Steidel} {et~al.}(1995){Steidel}, {Pettini}, \&
  {Hamilton}}]{steidel95}
{Steidel}, C.~C., {Pettini}, M., \& {Hamilton}, D. 1995, \aj, 110, 2519

\bibitem[{{Storrie-Lombardi} {et~al.}(1996){Storrie-Lombardi}, {McMahon}, \&
  {Irwin}}]{stolom96}
{Storrie-Lombardi}, L.~J., {McMahon}, R.~G., \& {Irwin}, M.~J. 1996, \mnras,
  283, L79

\bibitem[{{Storrie-Lombardi} \& {Wolfe}(2000)}]{stolom00}
{Storrie-Lombardi}, L.~J. \& {Wolfe}, A.~M. 2000, \apj, 543, 552

\bibitem[{{Taniguchi} \& {Shioya}(2001)}]{taniguchi01}
{Taniguchi}, Y. \& {Shioya}, Y. 2001, \apj, 547, 146

\bibitem[{{Urbaniak} \& {Wolfe}(1981)}]{urbaniak81}
{Urbaniak}, J.~J. \& {Wolfe}, A.~M. 1981, \apj, 244, 406

\bibitem[{{van Dokkum}(2001)}]{vandok01}
{van Dokkum}, P.~G. 2001, \pasp, 113, 1420

\bibitem[{{Warren} \& {M{\o}ller}(1996)}]{warren96}
{Warren}, S.~J. \& {M{\o}ller}, P. 1996, \aap, 311, 25

\bibitem[{{Warren} {et~al.}(2001){Warren}, {M{\o}ller}, {Fall}, \&
  {Jakobsen}}]{warren01}
{Warren}, S.~J., {M{\o}ller}, P., {Fall}, S.~M., \& {Jakobsen}, P. 2001,
  \mnras, 326, 759

\bibitem[{{Wolfe} {et~al.}(1995){Wolfe}, {Lanzetta}, {Foltz}, \&
  {Chaffee}}]{wolfe95}
{Wolfe}, A.~M., {Lanzetta}, K.~M., {Foltz}, C.~B., \& {Chaffee}, F.~H. 1995,
  \apj, 454, 698

\bibitem[{{Wolfe} {et~al.}(1986){Wolfe}, {Turnshek}, {Smith}, \&
  {Cohen}}]{wolfe86}
{Wolfe}, A.~M., {Turnshek}, D.~A., {Smith}, H.~E., \& {Cohen}, R.~D. 1986,
  \apjs, 61, 249

\bibitem[{{Zheng} \& {Miralda-Escud{\' e}}(2002)}]{zheng02}
{Zheng}, Z. \& {Miralda-Escud{\' e}}, J. 2002, \apj, 578, 33

\end{thebibliography}

\end{document}